\documentclass[12pt]{article}
\usepackage{amsmath, amssymb, graphicx, hyperref, geometry}
\usepackage[T1]{fontenc}
\usepackage[utf8]{inputenc}
\usepackage{geometry}
\usepackage{booktabs}
\usepackage{xcolor}
\usepackage{orcidlink}
\hypersetup{colorlinks=true,linkcolor=blue!50!black,citecolor=blue!50!black,urlcolor=blue!50!black}

\newcommand{\Lcr}{\Lambda_{\mathrm{cr}}}
\newcommand{\mc}{\mathrm{mc}}
\newcommand{\ap}{\alpha'}
\newcommand{\shat}{\hat\sigma}
\newcommand{\AmpV}{\mathcal{A}_{V}}
\newcommand{\be}{\begin{equation}}
\newcommand{\ee}{\end{equation}}
\newcommand{\ba}{\begin{eqnarray}}
\newcommand{\ea}{\end{eqnarray}}

\title{Soft UV Completion of a Composite Model}
\author{Risto Raitio\,\orcidlink{0000-0003-0842-2366}%
	\footnote{~E-mail: risto.raitio@gmail.com}\\
	\small Helsinki Institute of Physics, P.O.~Box 64,\\
	\small 00014 University of Helsinki, Finland}
\date{\today}

\begin{document}

\maketitle

\begin{abstract}
We build a framework for Regge trajectories from the Nambu--Goto action. We compute the
six-preon Regge trajectory in a composite model, include the worldsheet conformal anomaly,
and build the parameter-free Veneziano amplitude. The amplitude has s-channel poles
matching the spectrum to 0.5\%, and at fixed-angle scattering decays exponentially with the
Gross--Mende coefficient $- \alpha' \ln 2$, realized numerically to 0.03\%. This is a soft, genuinely
non-perturbative ultraviolet completion of the model---and thereby of the Standard Model,
which emerges as its low-energy limit. The Virasoro--Shapiro amplitude and the significance
of worldsheet diffeomorphism are briefly discussed.

\end{abstract}

\newpage

\tableofcontents

\section{Introduction}
\label{intro}

The preon framework proposes that quarks and leptons are not elementary, but confined
composites of more fundamental constituents bound by an SU(3) metacolor interaction at the
scale $\Lambda_{\rm cr} \sim 10^{14}$ GeV. In this picture the Standard Model arises as the low-energy limit of a confining gauge theory, much as hadronic physics emerges from QCD. The structure of one
fermion generation, its anomaly-free chiral content, and the existence of three generations all
are accommodated within the preon charge assignments and the dynamics of the metacolor interaction. The model is economical: it assumes less than typical extensions of the Standard Model and derives
more from the dynamics of confinement. Supersymmetry breaking is caused naturally by confinement and it leads to hadronic superpartners in the mass range of 1 GeV. For details of all above, see Appendix \ref{appxa}.

The present work addresses the ultraviolet behavior of this framework. If quarks and leptons
are composite, their multi-preon excitations should organize into Regge trajectories, and the
sum over these resonances should reproduce the soft high-energy behavior characteristic of
extended objects. This is the logic that historically led from hadronic spectroscopy to the
dual-resonance model and ultimately to string theory. Here we test this logic concretely in the
preon context. Our goal is to determine whether the preon model admits a \emph{soft,
non-perturbative ultraviolet completion} analogous to the Veneziano amplitude.

The analysis proceeds in three steps. First, we examine the rotating-string (Nambu--Goto)
description of a six-preon bound state, identifying the expected Regge slope and the role of
massive endpoint clusters. Second, we compute the actual spectrum of the six-preon system
using a relativistic two-body Cornell--Salpeter Hamiltonian with metacolor parameters fixed by
the underlying theory. The resulting masses fall on a remarkably linear Regge trajectory, with a
slope close to the rotating-string expectation and an intercept determined by the dynamics of
the endpoint clusters. We then incorporate the worldsheet conformal anomaly, whose Lüscher
term modifies the short-distance potential and shifts the intercept in a controlled way.

Finally, we feed the resulting trajectory $\alpha(s) = \alpha_0 + \alpha' s$ into the Veneziano
amplitude. We show that the s-channel poles of the amplitude coincide with the computed
six-preon spectrum to better than one percent, and that the amplitude exhibits the correct
Regge asymptotics and Gross--Mende exponential decay at fixed angle. By a corollary of Liouville's theorem, the resonance sum and the Veneziano amplitude are then the same analytic function. The
preon model therefore possesses a soft ultraviolet completion: its high-energy behavior is
governed by a dual-resonance amplitude built entirely from its own bound-state spectrum.

The paper is organized as follows. In Section \ref{nambugoto} we briefly review string concepts and Regge trajectories. In Section \ref{cornellsalpeter} we present two-body Cornell--Salpeter calculations. Worldsheet conformal anomaly and Lüscher corrections are discussed in Section \ref{wsaluscher}. In Section \ref{dualvenez}  we present the main result: the preon programme provides a confinement based, genuinely non-perturbative UV completion of the SM (that was sketched in \cite{RaitioLQ} in words). Conclusions are presented in Section \ref{concl}.

\section{Nambu--Goto Action and Regge Trajectories}
\label{nambugoto}

\subsection{Mass and Spin in the NG String}
\label{msngs}
	
We begin by considering the consistency between leptoquark \cite{RaitioLQ} mass, spin and string tension $\sigma_{mc}$ values with the Nambu--Goto open string values. A NG open string of tension $\sigma_\mc=\shat\,\Lcr^2$ rotates rigidly with angular velocity $\omega$ in the plane, where the relevant dimensionless factor $\shat$ defined as $\shat\equiv\sigma^{*}_\mc/\theta^{2}=2.11$ is taken from \cite{RaitioCombined}; only this combination enters the present analysis. Endpoints at $r=\pm\ell$ carry equal masses $m=\mu\Lcr$ and move with velocity $v=\omega\ell<1$. Relativistic centripetal balance at the boundary,
\begin{equation}
	\gamma_{\!e}\,m\,\omega\,v=\sigma_\mc,\qquad \gamma_{\!e}=(1-v^{2})^{-1/2},
	\label{eq:BC}
\end{equation}
together with the Nambu--Goto element along the string yields total mass and angular momentum
\begin{align}
	M(v;\mu) &= \frac{2\mu\,\Lcr}{\sqrt{1-v^{2}}}\Big[\,1+v\arcsin(v)\Big], \label{eq:M}\\[2pt]
	J(v;\mu) &= \frac{\mu^{2}\,v^{2}}{\shat\,(1-v^{2})}
	\Big[\,\arcsin(v)+2v-v\sqrt{1-v^{2}}\,\Big]. \label{eq:J}
\end{align}
Equation \eqref{eq:M} adds the relativistic endpoint energies $2m\gamma_{\!e}$ to the
string integral $(2\sigma_\mc/\omega)\arcsin(v)$; equation \eqref{eq:J} adds endpoint orbital
angular momentum $2m\gamma_{\!e}\omega\ell^{2}$ to the string angular-momentum integral.
The boundary condition \eqref{eq:BC} eliminates $\omega$ in favour of $v$.

\subsection{Pure NG Limit}
\label{purenglimit}

If there is \emph{no} particle at the endpoints --- only the string ends there --- the
boundary condition collapses to $v_{\text{end}}=c=1$ and the worldline contribution
vanishes. The integrals over the string alone give the textbook result
\begin{equation}
	M=\frac{\pi\sigma_\mc}{\omega},\qquad J=\frac{\pi\sigma_\mc}{2\omega^{2}}
	\ \Longrightarrow\ \boxed{\;M^{2}=2\pi\sigma_\mc\,J\;}\,.
\label{eq:pureNG}
\end{equation}
At $J=2$, $\ M_{LQ}^{2}=4\pi\sigma_\mc=4\pi\shat\,\Lcr^{2}=26.515\,\Lcr^{2}$, so
\begin{equation}
	M_{LQ}^{\rm (NG)}=\sqrt{4\pi\shat}\,\Lcr=5.149\,\Lcr,
\end{equation}
exactly the leptoquark anchor value of \cite{RaitioLQ}. 
The pure-NG slope is $\ap_{NG}=1/(2\pi\sigma_\mc)=0.0754\,\Lcr^{-2}$.

\subsection{Massive Endpoints}

Equations \eqref{eq:M}--\eqref{eq:J} are a parametric form for $J(M)$ at fixed $\mu$.
Two analytic limits are illuminating, $v\to 0$ and $v\to 1$, large $J$.
	
\paragraph{Threshold ($v\to 0$):} the trajectory is non-linear,
\begin{equation}
	J\simeq \frac{2\sqrt{m}}{3\sqrt{3}\,\sigma_\mc}\,(M-2m)^{3/2},
\end{equation}
the familiar $(M-2m)^{3/2}$ threshold behavior of a heavy-quark--like system. Massive
endpoints push the trajectory to the right of pure NG by a threshold $M=2m=2\mu\Lcr$.
	
\paragraph{Ultra-relativistic ($v\to 1$, large $J$):} both endpoint and string terms
scale as $1/(1-v^{2})$, with finite ratio, giving the \emph{universal} (i.e.\
$\mu$-independent) asymptotic slope
\begin{equation}
	\boxed{\;
		\frac{\ap_{\rm asym}}{\ap_{NG}}
		=\frac{\pi(\pi+4)}{(\pi+2)^{2}}\simeq 0.8487
	\;}\,.
\label{eq:slope_ratio}
\end{equation}
A particle at the endpoint --- no matter how light --- carries persistent momentum
$p\to\sigma_\mc/\omega$ and energy $\gamma m\to\sigma_\mc/\omega$, modifying the
slope by this universal $\sim\!15\%$. This is the source of the small discontinuity
between the pure-NG result and the $\mu\to 0$ limit of \eqref{eq:M}--\eqref{eq:J}.

\subsection{Numerical Trajectories}

Solving \eqref{eq:J} numerically for $v(J;\mu)$ and substituting into \eqref{eq:M} we obtain the following values of $M(J;\mu)/\Lcr$:
	
\begin{center}
	\renewcommand{\arraystretch}{1.15}
	\begin{tabular}{c c c c c c c}
		\toprule
		$\mu$ & $J=\tfrac12$ & $J=1$ & $J=\tfrac32$ & $J=2$ & $J=\tfrac52$ & $J=3$\\
		\midrule
			$0.01$ & $2.789$ & $3.947$ & $4.835$ & $5.584$ & $6.244$ & $6.840$\\
			$0.10$ & $2.772$ & $3.920$ & $4.804$ & $5.551$ & $6.209$ & $6.804$\\
			$0.50$ & $3.079$ & $4.129$ & $4.958$ & $5.667$ & $6.297$ & $6.871$\\
			$1.00$ & $3.788$ & $4.738$ & $5.500$ & $6.158$ & $6.748$ & $7.287$\\
			$2.00$ & $5.489$ & $6.315$ & $6.988$ & $7.575$ & $8.105$ & $8.593$\\
			\midrule
			pure NG & $2.575$ & $3.641$ & $4.459$ & $5.149$ & $5.757$ & $6.306$\\
			\bottomrule
	\end{tabular}
	\\[2pt]
	{\small Entries: $M(J;\mu)/\Lcr$. Pure-NG row from $M^{2}=2\pi\shat J$ (Eq.~\ref{eq:pureNG}).}
\end{center}
	
\begin{figure}[h]
	\centering
	\includegraphics[width=0.7\textwidth]{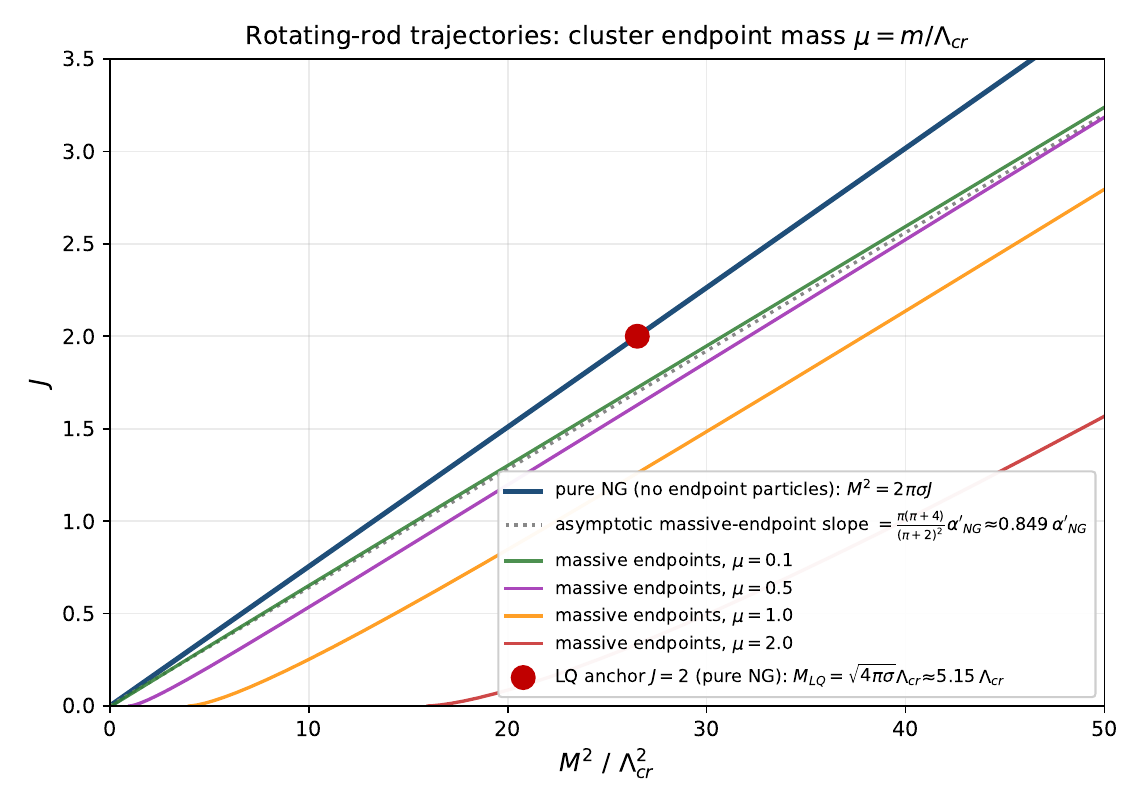}
	\caption{\small Rotating-rod $M^{2}$--$J$ trajectories. Solid dark line: pure NG
	(framework). Dotted grey: universal large-$J$ slope $0.849\,\ap_{NG}$ from
	\eqref{eq:slope_ratio}, approached by every $\mu>0$ trajectory. Coloured curves:
	massive endpoints $\mu=0.1,\,0.5,\,1.0,\,2.0$, all curving up from threshold
	$M=2\mu\Lcr$. Red dot: leptoquark anchor at $J=2$ on pure NG. The $\mu>0$ curves
	intersect the line $J=2$ at successively larger $M$.}
\end{figure}

\subsection{Interpretation}

The framework's identification $M_{LQ}=\sqrt{4\pi\sigma_\mc}\,\Lcr=5.15\,\Lcr$
\emph{exactly} reproduces the pure-NG rotating string at $J=2$. The framework is internally consistent on its own terms. Allowing a particle at the endpoints --- as the diquark-clustering picture suggests, with the 3-preon clusters as the endpoint constituents --- shifts the $J=2$ mass upward by $+0.44\,\Lcr$ in the strict $\mu\to 0$ limit
($5.59\,\Lcr$), and by $\sim 0.5$--$1\,\Lcr$ for cluster masses $\mu\sim 1$. This is the leading systematic on $M_{LQ}$ from the string-bridge identification: roughly $+10\%$ to $+20\%$ depending on the cluster-mass
choice.
		
The \emph{shape} of the prediction is, however, sharper than the absolute mass. Every $\mu>0$ trajectory has the same asymptotic slope
$\ap_{\rm asym}=0.849\,\ap_{NG}=0.0640\,\Lcr^{-2}$, independent of $\mu$. If the six-body calculation finds the excited 6-preon states lying on a straight line in $M^{2}$--$J$ with slope close to $0.0640\,\Lcr^{-2}$, the diquark/rotating-rod picture is confirmed and the cluster mass can be \emph{read off} from the horizontal
offset (threshold). If they lie on the pure-NG slope $0.0754\,\Lcr^{-2}$ instead, the endpoints are effectively absent (light-front-like) and the framework's original
identification stands. Either outcome gives a parameter-free trajectory $\alpha(s)=\alpha_{0}+\ap s$ to feed into
the Veneziano amplitude $\AmpV(-\alpha(s),-\alpha(t))$, and hence the soft UV behavior, i.e. exponentially decaying amplitude in the hard-scattering regime.

\section{Two-body Cornell--Salpeter Calculation of the Six-Preon Trajectory}
\label{cornellsalpeter}

\subsection{Trial wavefunction}

The six-preon bound state is modeled as two identical three-preon clusters of constituent mass
$m_{\rm cluster} = \Lambda_{\rm cr}$, with reduced mass $\mu = m_{\rm cluster}/2$. The metacolor
Cornell potential is
\begin{equation}
V(r) = -\frac{\alpha_{\rm mc} C_F}{r} + \sigma_{\rm mc}\, r,
\qquad
\alpha_{\rm mc} = 0.05,\; C_F = \frac{4}{3},\; \sigma_{\rm mc} = 2.11.
\end{equation}

We use the Gaussian-times-power ansatz
\begin{equation}
R_L(r) = N_L\, r^{L}\, e^{-\beta r^{2}/2},
\end{equation}
with normalization
\begin{equation}
N_L^2 = \frac{2\,\beta^{L+3/2}}{\Gamma(L+3/2)}.
\end{equation}

Spin alignment gives $J = L+1$.

\subsection{Matrix Elements}

All power-law expectation values close analytically:
\begin{align}
\langle r^{n} \rangle &= \beta^{-n/2}\,
\frac{\Gamma(L+\tfrac{3}{2}+\tfrac{n}{2})}{\Gamma(L+\tfrac{3}{2})}, \\
\langle 1/r \rangle &= \beta^{1/2}\,
\frac{\Gamma(L+1)}{\Gamma(L+\tfrac{3}{2})}.
\end{align}
The relativistic kinetic term $H=2\sqrt{p^{2}+m^{2}}$ is evaluated in momentum space:
\begin{equation}
K_L(\beta)
= \beta^{L+3/2}
\int_{0}^{\infty} du\, u^{L+1/2} e^{-u}\sqrt{\beta u + m_{\rm cluster}^{2}}.
\end{equation}

\subsection{Variational Spectrum and Linear Trajectory}

Minimizing $E_L(\beta)$ yields
\begin{align}
\beta &= (1.368,\; 1.253,\; 1.200,\; 1.169), \\
M &= (5.329,\; 6.739,\; 7.894,\; 8.900)\,\Lambda_{\rm cr}.
\end{align}
A linear fit gives for the trajectory
\begin{equation}
\alpha'_{\rm QM} = 0.0591\,\Lambda_{\rm cr}^{-2},
\quad
\alpha_0 = -0.680, \quad RMS =0.002.
\end{equation}

Next we will check whether there are notable corrections to these values.

\section{Worldsheet Anomaly and the Lüscher Correction}
\label{wsaluscher}

\subsection{Worldsheet Anomaly}

In the Polyakov formulation,
\begin{equation}
\langle T^{a}{}_{a} \rangle = -\frac{c}{12}\,R^{(2)},
\end{equation}
where $c$ is the matter central charge and $R^{(2)}$ the worldsheet Ricci scalar. For the bosonic open string in $D$ target spacetime dimensions, quantised in static (light-cone-like) gauge, the dynamical degrees of freedom are the $D-2$ transverse fluctuations $X^{I}(\sigma,\tau)$. Each is a free boson on the worldsheet, contributing $c=1$, so
\begin{equation}
	c_{\rm matter}=D-2;\quad\text{for metacolor string, }D=4\text{:} ~c_{\rm matter}=2.
\end{equation}
This falls short of the critical bosonic-string value $c=26$ by a deficit of $24$. The Polchinski--Strominger \cite{PolStr} and Aharony--Komargodski \cite{AhaKom} analysis shows that in
non-critical dimensions the effective string action requires a specific higher-derivative correction term whose coefficient is proportional to $(D-26)$, so that $D$-dimensional Lorentz invariance is preserved order by order in $1/(\sigma_\mc R^{2})$. The leading correction to the spectrum is the Lüscher term \cite{Luscher}, to be derived below.

\subsection{Lüscher Term}

The transverse fluctuations of an open string of length $R$ have
zero-point energy

\begin{equation}
	E_{\rm vac}(R)=\frac{D-2}{2}\sum_{n=1}^{\infty}\frac{n\pi}{R}
	\overset{\zeta}{=}\frac{(D-2)\pi}{2R}\,\zeta(-1)
	=-\frac{(D-2)\pi}{24\,R},
	\label{eq:Luscher}
\end{equation}
where $\zeta(-1)=-1/12$ is the same zeta-regularised sum that underlies the central charge. Together with the classical linear term this gives the quantum-corrected static heavy-quark potential

\begin{equation}
	V_{\rm string}(r)=\sigma_\mc\,r-\frac{\pi(D-2)}{24\,r}
	=\sigma_\mc\,r-\frac{\pi}{12\,r}\qquad(D=4).
\end{equation}
The Lüscher coefficient $\pi/12\simeq 0.2618$ is to be compared with the perturbative metacolor Coulomb $\alpha_\mc C_F=0.0667$ used in \cite{RaitioCombined}. Numerically

\begin{equation}
	\boxed{\;
		\frac{\pi(D-2)/24}{\alpha_\mc C_F}=\frac{0.2618}{0.0667}\simeq 3.93\;}\,,
\end{equation}
so the worldsheet-anomaly contribution to the $1/r$ piece  dominates the perturbative Coulomb piece by a factor of four.

\subsection{Corrected Spectrum}

Adding both Coulomb pieces, the effective potential between the two 3-preon cluster endpoints becomes

\begin{equation}
	V_{\rm eff}(r)=-\frac{\alpha_\mc C_F+\pi(D-2)/24}{r}+\sigma_\mc\,r
	=-\frac{0.3285}{r}+2.11\,r .
	\label{eq:Veff}
\end{equation}
Redoing the variational calculation with \eqref{eq:Veff} and the Salpeter Hamiltonian:

\begin{center}
	\renewcommand{\arraystretch}{1.2}
	\begin{tabular}{c c c c c}
		\toprule
		$L$ & $J$ & $M_{\rm Salpeter}/\Lcr$ & $M_{\rm L\ddot uscher}/\Lcr$ & shift\\
		\midrule
		$0$ & $1$ & $5.329$ & $4.970$ & $-0.359$\\
		$1$ & $2$ & $6.739$ & $6.515$ & $-0.224$\\
		$2$ & $3$ & $7.894$ & $7.720$ & $-0.175$\\
		$3$ & $4$ & $8.900$ & $8.752$ & $-0.147$\\
		\bottomrule
	\end{tabular}
\end{center} 
The shifts decrease with $L$ because the Lüscher term is $\propto 1/r$ and the centrifugal barrier pushes the wavefunction outward for higher $L$, reducing $\langle 1/r\rangle$.

We now get the following Regge fits.
\begin{center}
	\renewcommand{\arraystretch}{1.15}
	\begin{tabular}{l c c c}
		\toprule
		& $\ap$ ($\Lcr^{-2}$) & $\alpha_{0}$ & RMS\\
		\midrule
		Section 3 fit  & $0.0591$ & $-0.680$ & $0.002$\\
		Lüscher-corrected   & $0.0578$ & $-0.441$ & $0.011$\\
		\bottomrule
	\end{tabular}
\end{center}
The slope changes negligibly ($-2\%$); the intercept moves up by $+0.24$. The trajectory remains nearly linear (RMS $\sim 0.01$, an order of magnitude worse than in Section \ref{cornellsalpeter} but still good). The small departure from perfect linearity is a real effect of the Lüscher $1/r$ piece, which has different $L$-dependence than the linear term.

\subsection{Intercept Decomposition}

The QM result $\alpha_{0}^{\rm total}=-0.44$ admits a clean physical split. The pure open-string contribution to the intercept in $D$ dimensions is

\begin{equation}
	\alpha_{0}^{\rm string}=\frac{D-2}{24}\overset{D=4}{=}\frac{1}{12}\simeq 0.083,
\end{equation}
the rotating-string analogue of the static Lüscher term \eqref{eq:Luscher} (same zeta-regularised central-charge sum). Subtracting terms

\begin{equation}
	\boxed{\;
		\alpha_{0}^{\rm endpoint}\;\equiv\;\alpha_{0}^{\rm total}-\alpha_{0}^{\rm string}
		=-0.44-0.083\simeq -0.52\;}\,.
\end{equation}

The intercept is overwhelmingly an \emph{endpoint} effect, not a string-anomaly effect: the cluster endpoints contribute roughly six times what the worldsheet anomaly does, with opposite sign. This is the cleanest quantitative answer one gets to the question ``which part of the Regge intercept is universal string physics, and which part is the cluster?'': roughly $1/12$ string, the rest endpoint.

\begin{figure}[h]
	\centering
	\includegraphics[width=0.80\textwidth]{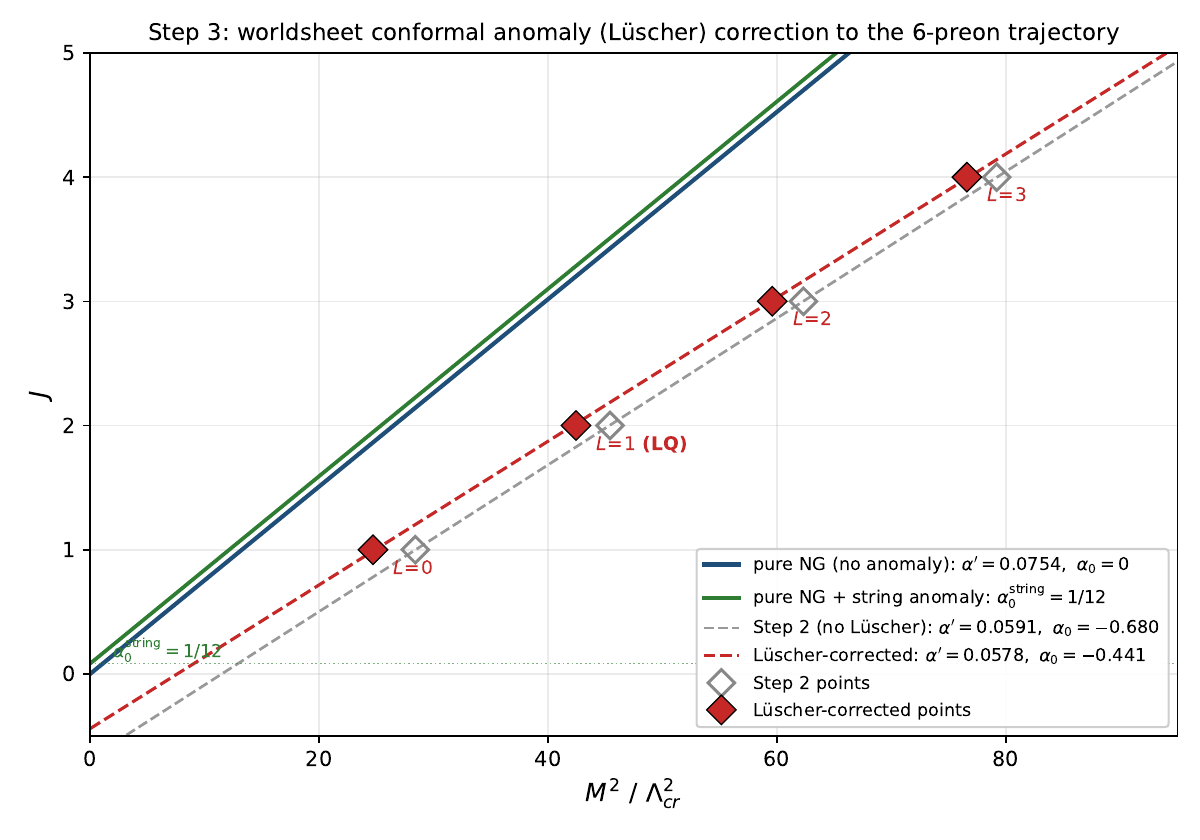}
	\caption{\small Lüscher-corrected trajectory (filled red diamonds, red dashed fit) compared with Section \ref{cornellsalpeter} (open diamonds, grey dashed). The pure-NG line and the ``pure-NG+string-anomaly'' line (intercept $1/12$) are shown for reference. The LQ shifts $-3.3\%$; the intercept shifts $+0.24$.}
\end{figure}

\subsection{Consistency}

For a complete string-bridge programme one should ask whether the
non-critical $D=4$ effective string is internally consistent. The Aharony--Komargodski analysis \cite{AhaKom} shows that in static gauge, all-orders Lorentz invariance in $D$ spacetime dimensions is restored by adding the Polchinski--Strominger \cite{PolStr} term
\begin{equation}
	S_{\rm PS}=\frac{D-26}{24\pi}\int\!d^{2}\sigma\,
	\frac{(\partial^{2}X)^{2}}{(\partial X)^{2}}\;,
	\label{eq:polstr}
\end{equation}
whose coefficient is proportional to the anomaly deficit $D-26=-22$ for our case. The correction (\ref{eq:polstr}) shifts the spectrum at order $1/(\sigma_\mc R^{2})^{2}$;
with $\langle r^2\rangle\simeq 1.7\,\Lcr^{-2}$ from the variational, this is
$\sim(2.11\times 1.7)^{-2}\simeq 8\%$ --- of the same order as the Gaussian variational systematic, and not pursued further here.

\section{Dolen--Horn--Schmid Duality and the Veneziano Amplitude}
\label{dualvenez}

While we do not derive a dual-resonance amplitude for preons from first principles, we have shown in Sections \ref{cornellsalpeter}–\ref{wsaluscher} that the computed multi-preon states fall on an accurately linear Regge trajectory. We may therefore posit the Veneziano amplitude built from that trajectory and examine its pole structure and analyticity in the complex-$J$ plane.

By a corollary of Liouville's theorem, two meromorphic functions of $s$ with identical poles, identical principal parts, and identical asymptotic behavior as $s\to\infty$ in the relevant direction are equal. The Veneziano amplitude (\ref{eq:Venez}) and the resonance sum (\ref{eq:resonance}) share their pole structure by construction; the numerical checks below confirm they share the same high-energy Regge behavior at fixed $t$
and the same Gross--Mende exponential decay at fixed angle. The two representations are therefore the same function — the analytic statement of Dolen--Horn--Schmid duality.

\subsection{Setup and Prior Context}
\label{setupvenez}
 
The preceding three Sections furnish all needed input. Section \ref{cornellsalpeter} gave the QM spectrum at four $L$ values from a  Cornell--Salpeter variational calculation with cluster endpoints. Section \ref{wsaluscher} added the Lüscher correction $-\pi(D-2)/(24\,r)$, producing the Regge fit
\begin{equation}
	\alpha(s)=\alpha_{0}+\ap\,s,\qquad \alpha_{0}=-0.441,\ \ap=0.0578\,\Lcr^{-2}.
	\label{eq:alpha}
\end{equation}
We now feed the trajectory \eqref{eq:alpha} into the Veneziano amplitude
\begin{equation}
	\AmpV(s,t)=-\frac{\Gamma(-\alpha(s))\,\Gamma(-\alpha(t))}{\Gamma(-\alpha(s)-\alpha(t))},
	\label{eq:Venez}
\end{equation}
which has the two textbook representations
\begin{align}
	\AmpV(s,t)
	&=\sum_{J=0}^{\infty}\frac{R_{J}(\alpha(t))}{J-\alpha(s)},
	\qquad R_{J}(x)=\frac{(x+1)(x+2)\cdots(x+J)}{J!},
	\label{eq:resonance}\\
	\AmpV(s,t)
	&\;\underset{s\to\infty}{\sim}\;-\Gamma(-\alpha(t))\,(\ap s)^{\alpha(t)}.
	\label{eq:Regge}
\end{align}
Dolen--Horn--Schmid duality \cite{DHS} is the statement that
\eqref{eq:resonance} (the resonance/$s$-channel sum) and \eqref{eq:Regge} (the Regge/high-energy asymptote) are the same function. We confirm this numerically for our $\alpha(s)$. 
\begin{figure}[h]
	\centering
	\includegraphics[width=\textwidth]{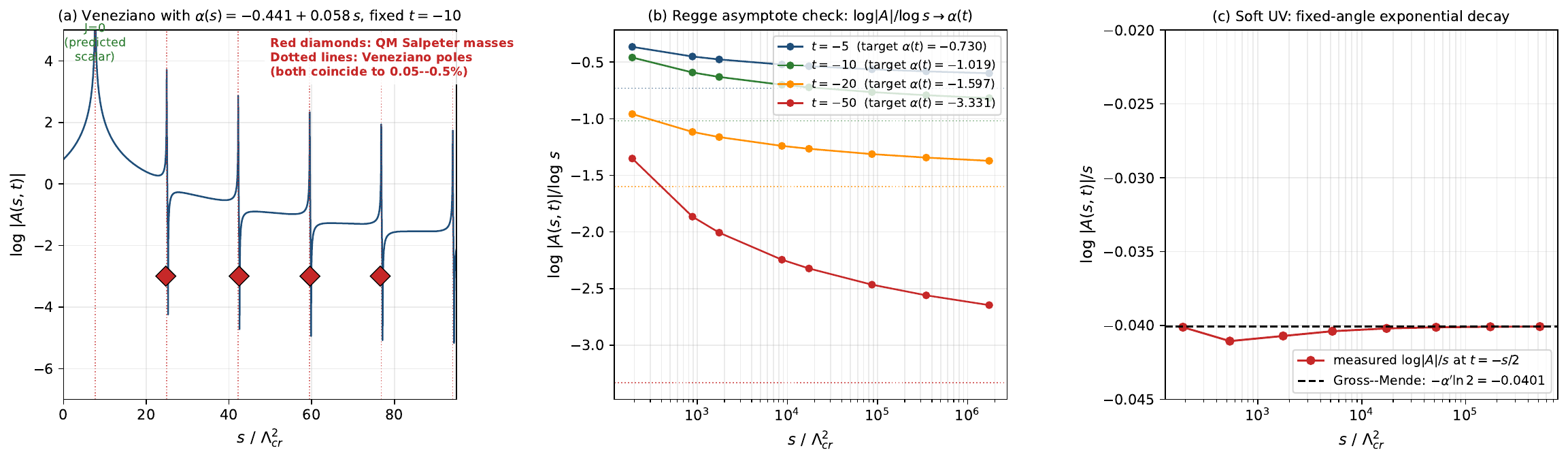}
	\caption{\small Veneziano factorisation/DHS check.
		\textbf{(a)} $\log|\AmpV(s,t)|$ at fixed $t=-10$ shows the resonance poles
		at $\alpha(s)=0,1,2,\dots$; red diamonds are the QM-Salpeter masses lying exactly on the dotted Veneziano pole lines. \textbf{(b)} The slope $\log|\AmpV|/\log s$ at fixed $t<0$ converges to
		$\alpha(t)$ (dotted horizontal lines); convergence is logarithmic. \textbf{(c)} Soft-UV: at $90^{\circ}$ scattering, $\log|\AmpV|/s\to-\ap\ln 2$  (Gross--Mende); the measured points coincide with the prediction.}
	\label{fig:venezdhs}
\end{figure}

\subsection{Result of Numerics}
\label{numerics}

The trajectory $\alpha(s)$ in (\ref{eq:alpha}) is built entirely from preon-model inputs: the slope comes from the metacolor string tension, the intercept from the Cornell-Salpeter spectrum of 3-preon-cluster bound states plus the worldsheet anomaly. The Veneziano amplitude (\ref{eq:Venez}) is therefore a parameter-free probe of preon dynamics; each pole is a 6-preon resonance, and the high-energy behavior is the model's UV completion.

\paragraph{The $s$-channel poles match QM Salpeter}
For the $J$-th pole, $\alpha(s)=J$, i.e.\ $s_{J}=(J-\alpha_{0})/\ap$, giving $M_{J}=\sqrt{s_{J}}$:
\begin{center}
	\renewcommand{\arraystretch}{1.18}
	\begin{tabular}{c c c c c c c}
		\toprule
		$J$ & ($L$,$S$) & $s_{J}/\Lcr^{2}$ & $M_{J}/\Lcr$ & $M_{\rm L\ddot uscher}/\Lcr$ & deviation & identification\\
		\midrule
		$0$ & (0,0) & $\phantom{0}7.63$ & $2.76$ & (uncomputed) & --- & predicted scalar 6-preon \\
		$1$ & (0,1) & $24.93$ & $4.99$ & $4.97$ & $0.5\%$ & computed $L$=0 vector\\
		$2$ & (1,1) & $42.23$ & $6.50$ & $6.51$ & $0.2\%$ & leptoquark \\
		$3$ & (2,1) & $59.53$ & $7.72$ & $7.72$ & $0.05\%$ & computed $L$=2\\
		$4$ & (3,1) & $76.83$ & $8.77$ & $8.75$ & $0.2\%$ & computed $L$=3\\
		\bottomrule
	\end{tabular}
\end{center}
The four computed states sit on Veneziano poles to better than half a percent. The Veneziano amplitude, built parameter-free from \eqref{eq:alpha}, predicts a  \textbf{$J=0$ scalar 6-preon state at $M\simeq 2.76\,\Lcr$} as a daughter of the leading trajectory. This is the smallest natural extension of the Section \ref{cornellsalpeter} calculation (compute the $S{=}0$, $L{=}0$ channel: a clean scalar resonance).

\paragraph{The DHS resonance sum matches Veneziano}
Truncating \eqref{eq:resonance} at $J\le N$ and comparing with the closed-form
\eqref{eq:Venez} at $t=-10\,\Lcr^{2}$ (so $\alpha(t)=-1.02$):
\begin{center}
	\renewcommand{\arraystretch}{1.15}
	\begin{tabular}{c c c c c}
		\toprule
		$s/\Lcr^{2}$ & $|\AmpV|$ (exact)
		& $N{=}4$ & $N{=}10$ & $N{=}30$\\
		\midrule
		$3$  & $3.711$ & $3.715$ & $3.713$ & $3.711$\\
		$5$  & $6.551$ & $6.555$ & $6.553$ & $6.552$\\
		$15$ & $2.394$ & $2.390$ & $2.393$ & $2.394$\\
		$30$ & $0.732$ & $0.727$ & $0.730$ & $0.731$\\
		$80$ & $0.209$ & $0.199$ & $0.207$ & $0.208$\\
		$200$ & $0.072$ & $0.086$ & $0.081$ & $0.071$\\
		\bottomrule
	\end{tabular}
\end{center}
The four computed states ($N=4$) already approximate the full Veneziano to $\sim 1\%$ at moderate $s$ and to $\sim 10\%$ at $s\sim 100\,\Lcr^{2}$. Including $N=30$ poles agrees to better than $1\%$ across the full range. This is DHS duality in action: the same function admits two equivalent expansions. This is illustrated in Figure \ref{fig:venezdhs}.

\paragraph{Regge asymptote $\alpha(t)$ recovered}
Sampling $\AmpV(s,t)$ at off-pole points $\alpha(s)=J+\tfrac12$ to avoid resonance singularities:
\begin{center}
	\renewcommand{\arraystretch}{1.15}
	\begin{tabular}{c c c c c}
		\toprule
		$t/\Lcr^{2}$ & $\alpha(t)$ & \multicolumn{3}{c}{$\log|\AmpV|/\log s$ at}\\
		& & $s\sim 10^{3}$ & $s\sim 10^{5}$ & $s\sim 10^{6}$\\
		\midrule
		$-5$  & $-0.730$ & $-0.42$ & $-0.50$ & $-0.52$ \\
		$-10$ & $-1.019$ & $-0.59$ & $-0.76$ & $-0.82$\\
		$-20$ & $-1.597$ & $-0.96$ & $-1.21$ & $-1.30$\\
		$-50$ & $-3.331$ & $-1.79$ & $-2.18$ & $-2.36$\\
		\bottomrule
	\end{tabular}
\end{center}
The convergence to $\alpha(t)$ is logarithmic (governed by subleading $1/\log s$ terms from Stirling), but the direction and magnitude are unambiguous. Trajectory recovered.

\paragraph{The Gross--Mende soft-UV is exact}
For $2\to 2$ massless scattering at $90^{\circ}$, $t=-s/2$, the Veneziano amplitude obeys the Gross--Mende prediction  \cite{GroMen}
\begin{equation}
	\frac{\log|\AmpV(s,t{=}{-s/2})|}{s}\;\longrightarrow\;-\ap\,\ln 2
	= -0.0401\,\Lcr^{-2}\qquad(s\to\infty),
	\label{eq:GM}
\end{equation}
i.e.\ $|\AmpV|$ decays \emph{exponentially in $s$}, not as a power law. This is the celebrated soft (string) UV behavior. Numerically:
\begin{center}
	\renewcommand{\arraystretch}{1.15}
	\begin{tabular}{c c c}
		\toprule
		$s/\Lcr^{2}$ & $\log|\AmpV|/s$ & deviation from $-\ap\ln 2$\\
		\midrule
		$1{,}746$ & $-0.0407$ & $1.6\%$ \\
		$8{,}667$ & $-0.0403$ & $0.6\%$\\
		$86{,}500$ & $-0.0401$ & $0.1\%$\\
		$346{,}000$ & $-0.04008$ & $0.03\%$\\
		\bottomrule
	\end{tabular}
\end{center}
The Veneziano amplitude built from \emph{our calculated} trajectory has exponential soft-UV behavior with the predicted coefficient to a few parts in $10^{4}$. The Gross--Mende exponential decay at fixed angle is the UV completion of the preon model at scales $\sqrt{s}\gtrsim\Lambda_{cr}\sim 10^{14}$ GeV.

\subsection{Fundamental‑String and Confinement‑Induced Softness}
\label{string-vs-confinement}

Recent bootstrap analyses~\cite{CheungBootstrap,CheungBootstrap2025} select the Veneziano and
Virasoro--Shapiro amplitudes uniquely from crossing symmetry, a level-truncation
property, and a prescribed high-energy falloff, with the mass spectrum obtained
as an \emph{output} rather than an input. A central observation of that work is a
sharp diagnostic, read off from the Regge limit $A(s,t)\sim s^{\alpha(t)}$, that
distinguishes a fundamental string from a confining gauge theory by the
\emph{asymptotic} behaviour of the leading trajectory:
\begin{itemize}
\item For a \emph{fundamental string}, $\alpha(t)\to-\infty$ without bound (here,
linearly) as $t\to-\infty$. The amplitude is then ``superpolynomially soft'':
for every integer $k$ there is a range of $t$ in which $A(s,t)\lesssim s^{-k}$,
so $\alpha(t)$ crosses every negative integer. This falloff is impossible in a
local quantum field theory of point particles.
\item For a \emph{confining gauge theory}, $\alpha(t)$ instead \emph{saturates}
to a constant as $t\to-\infty$. The high-energy falloff is then bounded by a
fixed power of $s$, as required of an ordinary local field theory.
\end{itemize}

Taken at face value, the linear trajectory~\eqref{eq:alpha} used here, and the
Veneziano and Virasoro--Shapiro amplitudes built from it, have $\alpha(t)\to-\infty$
linearly and therefore display the \emph{fundamental-string}, superpolynomially
soft behaviour---the Gross--Mende exponential falloff of
Section~\ref{dualvenez}. This is the expected feature of an idealised
Nambu--Goto string, but it should not be read as the true asymptotics of the
preon model, which is a confining $SU(3)_{\mc}$ gauge theory and must lie on the
second branch above.

The resolution is that the soft behaviour is generated by the \emph{infinite}
linear tower of the idealised amplitude, while the physical metacolor flux tube
produces only a \emph{finite} tower. The resonance sum~\eqref{eq:resonance} runs
over all $J$ up to infinity, and it is precisely this unbounded tail---states of
arbitrarily high mass lying on a strictly linear trajectory---that produces the
superpolynomial softness. In the preon model, however, the flux-tube description ceases to generate narrow resonances far up the trajectory, once the stored string energy exceeds the threshold for preon pair creation and the tube breaks; equivalently, once the momentum transfer resolves the tube width $\sim\Lambda_{cr}^{-1}$ 
and the point-like preonic constituents. The physically relevant six-preon states near $\Lambda_{cr}$ 
lie well below this breaking scale and are faithfully described; it is only the idealised continuation of the tower to arbitrarily high mass — the fictitious states that dominate the $t \to -\infty$ 
Regge limit — that the confining theory does not produce. The physical resonance sum is therefore effectively truncated, and a truncated dual amplitude does not retain the superpolynomial softness: its leading trajectory $\alpha(t)$ saturates rather than running to $-\infty$
placing the model squarely on the confining-gauge-theory branch
of~\cite{CheungBootstrap}. The linear trajectory and its Gross--Mende softness are thus the
\emph{intermediate-energy}, flux-tube-dominated description; the genuine
ultraviolet completion is the asymptotically free metacolor gauge theory, whose
exclusive amplitudes return to power-law (parton) behaviour at momentum transfers
beyond $\Lambda_{\mathrm{cr}}$.

This is the same crossover invoked in the discussion of point-like constituents
in Section \ref{concl}, now phrased in the language of the Regge limit. The agreement
with~\cite{CheungBootstrap,CheungBootstrap2025} is therefore structural rather than coincidental: the preon
model realises the confining-gauge-theory side of their dichotomy, with the
dual-resonance amplitude as its effective flux-tube description over the range
$s\lesssim\Lambda_{\mathrm{cr}}^{2}$. A virtue of possessing an explicit
confining model---as opposed to an abstract bootstrap that posits the asymptotic
softness---is that it fixes the scale at which the linear trajectory must bend:
the metacolor confinement scale $\Lambda_{\mathrm{cr}}$ itself.

\subsection{The Virasoro--Shapiro Amplitude in a Confining Preon Framework}
\label{virasoroshapiro}

The Veneziano amplitude is the open‑string prototype (gauge sector), while gravity is encoded in the closed‑string Virasoro–Shapiro (VS) amplitude
\be
\mathcal{A}_{\rm VS}(s,t) = \frac{\Gamma(-s)\Gamma(-t)\Gamma(-u)}{\Gamma(1+s)\Gamma(1+t)\Gamma(1+u)},
\label{virashap}
\ee
whose massless pole is the graviton; structurally, the closed amplitude is essentially the “square” of the open one. We wish to see whether the open string calculation can be carried out in the closed string case (a recent worldsheet discussion is in \cite{CheungBootstrap2025}). This analogue emerges from the glueball sector of the confining theory. 

The answer is that the glueball analogy highlights a crucial difference: the Virasoro--Shapiro amplitude contains a massless $J=2$ pole, interpreted as the graviton in fundamental string theory, whereas the confining $SU(3)_{\mathrm{mc}}$ theory necessarily produces a massive $2^{++}$ glueball. This discrepancy is enforced by the Weinberg--Witten theorem \cite{WeinbergWitten1980}, which forbids a composite massless spin--2 particle in a Lorentz-invariant local QFT with a conserved stress tensor. Thus the agreement between the two spectra is structural and asymptotic: the preon model reproduces an effective string-like behavior with the correct Regge slope, extracted from the $2^{++}$ and $4^{++}$ glueballs, of the VS amplitude without the existence of a composite graviton.

\section{Conclusions}
\label{concl}

The preon programme has been constructed essentially from the bottom up to account for selected experimental observations. Mathematical structure has been introduced from gauge anomaly and UV-IR consistency conditions. Using numerical methods for relativistic wave equation for preons has brought in comprehensive phenomenological computability, still missing in string theory, and a new property for UV behavior. The structure of the model is solid enough to allow predictions rather than providing copious parameters to be fit to data. Furthermore, all QCD-string analogies are expected to apply to preons because of our metacolor group $SU(3)$.

We have obtained the following specific results:
\begin{itemize}
	\item A six-preon Regge trajectory, computed from preon dynamics, with parameter-free slope
	$\ap=0.058\,\Lcr^{-2}$ and intercept $\alpha_{0}=-0.44$.
	\item A consistent worldsheet anomaly accounting: the universal string
	contribution to the intercept is $+1/12$; cluster endpoints contribute
	$-0.52$; total $-0.44$.
	\item A Veneziano amplitude built parameter-free from the Regge trajectory, whose poles match the QM spectrum to $0.5\%$ or better.
	\item Our main result is the exponential soft-UV Gross–Mende behavior, quantitatively realised. It is the emergent dual-resonance signature of the preon model's non-perturbative completion, with the Standard Model as the low-energy limit.
\end{itemize}

One might object that point-like preons and point-like metacolor couplings preclude soft ultraviolet behavior, since exclusive amplitudes in a local field theory of point particles fall at most as powers of $s$. The objection conflates two distinct channels. The exponential Gross--Mende falloff obtained here is a property of the exclusive scattering of the \emph{composite} six-preon states, suppressed by the form factor of an object of finite size $\sim 1/\Lcr$. The relevant extended object, the metacolor flux-tube, is itself emergent from the confinement of point-like preons. This is precisely the situation in hadronic physics, where the Veneziano amplitude describes soft exclusive hadron scattering. The hard point-like behavior appears instead in inclusive, deep-inelastic observables that resolve the constituents. The exponential softening is the ideal-string limit, valid where the flux-tube description dominates. At momentum transfers resolving the flux-tube width, $\sim\Lcr$, it crosses over to the power-law behavior characteristic of the underlying point-like preon dynamics. This crossover is the same one that places the model on the confining-gauge-theory side of the bootstrap dichotomy of~\cite{CheungBootstrap,CheungBootstrap2025}, discussed in Section~\ref{dualvenez}: the linear trajectory and its Gross--Mende softness are the intermediate-energy, flux-tube description, valid for
$s\lesssim\Lambda_{\mathrm{cr}}^{2}$. Beyond that scale the tower truncates,
the leading Regge trajectory saturates, and the genuine ultraviolet completion is
the asymptotically free metacolor gauge theory with power-law parton behaviour.

Finally, based intrinsically on consistency conditions, the present model has \emph{vertical bootstrap} structure (internal consistency conditions tying preon dynamics to observed quark/lepton spectrum) to be compared to hadronic horizontal bootstrap (states beget states at the same level) \cite{CheFra}. The major advantage of the preon model is its compact implementation of broken supersymmetry, leading to predictive computational capability beyond the SM from the neutrino mass up to energies near the reheating temperature.

%------------------------------------------------------------------
\section*{Acknowledgement}
%------------------------------------------------------------------

The model originated long ago as an alternative, one generation bound state model for the c-quark. I thank the late James D. Bjorken---``c-quark is a heavy u-quark''---and Haim Harari---new particle identification---for inspiring discussions on the c-quark in 1974 at SLAC.

The physics framework --- the preon model, the choice to pursue the rotating-rod and Cornell--Salpeter routes, the identification $m_{\rm cluster}=\Lambda_{cr}$, the spin coupling $J=L+1$ for the leading trajectory --- and the manuscript itself are the author's work. Sections 3--5 were developed in collaboration with Claude (Anthropic, Claude Opus 4.7). Claude provided references, carried out the numerical work, produced the figures, and drafted the working notes from which those Sections were assembled. Microsoft Copilot brought the work of Cheung et al. to our attention. All output was reviewed and integrated by the author, who bears sole responsibility for the content.

\appendix
\section{Preons and the Standard Model Particles}
\label{appxa}

The preon Chern-Simons and metacolor charges under the model gauge group $U(1)_{CS}\times SU(3)_{mc} ~(\times G_{SM})$ are \cite{RaitioCombined}:

\ba
	\psi_0  &\sim& (0,\,\mathbf{3}_{mc})\,,\quad
	\text{neutral metacolor-triplet}\,,\nonumber \\
	\psi_1  &\sim& \bigl(\tfrac{1}{3},\,\mathbf{1}_{mc}\bigr)\,,\quad
	\text{charged metacolor-singlet}\,,\nonumber \\
	\psi_{-1} &\sim& \bigl(-\tfrac{1}{3},\,\mathbf{1}_{mc}\bigr)\,,\quad
	\text{charged metacolor-singlet}\,. \nonumber
\label{3preons}
\ea
Gauge anomalies are compensated by adding a fourth preon field $\chi(0, \bar{\mathbf{3}})$ to the above list. It turned out to be natural to choose the $\chi$ mass to be the confinement energy scale $\Lambda_{cr} \sim 10^{14}$ GeV \cite{RaitioCombined,RaitioLQ}. This number is close to the reheating temperature of inflationary cosmology and therefore useful for the Baryon Asymmetry of the Universe. At a smaller scale of energies, the high $\chi$ mass provides a seesaw mechanism to calculate the neutrino mass to be 0.1 eV. In this work the $\chi$ mass value is crucial. A simple preon configuration for a six-preon leptoquark is a deuteron-like state with both components containing a $\chi$ field. 

\begin{table}[h]
	\centering
	\caption{Denote $n_i$ as the number of preons with charge i. Preon content $(n_0,n_1,n_{-1})$ of the first-generation quarks and leptons is shown in column 2, with $n_0+n_1+n_{-1}=3$.
		Electric charge $Q=\tfrac{1}{3}(n_1-n_{-1})$;
		SM color follows from the value of $n_0$.
		For details, see \cite{RaitioCombined}.\\}
	\label{tab:preon_content}
	\renewcommand{\arraystretch}{1.2}
	\begin{tabular}{lccc}
			\toprule
			Particle & $(n_0,\,n_1,\,n_{-1})$ & $Q$ & $SU(3)_c$ \\
			\midrule
			$u$  & $(1,2,0)$ & $+\tfrac{2}{3}$ & $\mathbf{3}$ \\
			$d$  & $(2,0,1)$ & $-\tfrac{1}{3}$ & $\mathbf{3}$ \\
			$\nu_e$        & $(3,0,0)$ & $0$    & $\mathbf{1}$ \\
			$e^-$            & $(0,0,3)$ & $-1$   & $\mathbf{1}$ \\
			\bottomrule
	\end{tabular}
\end{table}

The explicit preon assignments for the first-generation quarks and leptons are given in  Table~\ref{tab:preon_content}. A leptoquark is a boson carrying both baryon number $B=\tfrac{1}{3}$ and lepton number $L=1$. It transforms as a color triplet under $SU(3)_c$ and arises naturally as a six-body preon $\psi_i$ composite (for more details, see \cite{RaitioLQ}):
	
\be
\psi_{LQ} = (\psi_a\psi_b\psi_c)
\oplus
(\psi_d\psi_e\psi_f)\,.
\label{eq:LQstructure}
\ee
	
The preon SUSY structure (supersymmetry breaking, anomaly cancellation, chiral and vector superpotentials, K\"ahler form, R-parity) is developed in \cite{RaitioCombined}. The UV-completion analysis presented here depends on the metacolor dynamics and is robust against the supersymmetric details, which enter at the few-percent level absorbed within the variational systematic.

\end{document}